\documentclass[camera,letterpaper,nomarginnotes,nonarrowgutter]{jpaper} 

\usepackage[sort,compress]{cite}
\usepackage{amsmath,amssymb,amsfonts}
\usepackage{algorithmic}
\usepackage{graphicx}
\usepackage{textcomp}
\usepackage{dblfloatfix} 
\usepackage{balance}
\usepackage{setspace}
\usepackage{xcolor}
\usepackage[bookmarks=true,breaklinks=true,letterpaper=true,colorlinks,linkcolor=red,citecolor=blue,urlcolor=magenta]{hyperref}
\usepackage{pifont}

\usepackage{fancyhdr}
\usepackage{xspace}
\usepackage[normalem]{ulem}
\usepackage{booktabs}
\usepackage{siunitx}
\usepackage{soul}
\usepackage{setspace}

\newcommand{\rev}[1]{\textcolor{black}{#1}}
\newcommand{\revt}[1]{\textcolor{black}{#1}}
\newcommand{\revz}[1]{\textcolor{black}{#1}}
\newcommand{\revd}[1]{\textcolor{black}{#1}}
\newcommand{\revv}[1]{\textcolor{black}{#1}}

\newcommand{\affilETH}[0]{\textsuperscript{\S}}
\newcommand{\affilETU}[0]{\textsuperscript{$\dagger$}}

\hypersetup{colorlinks,urlcolor=blue}
\def\BibTeX{{\rm B\kern-.05em{\sc i\kern-.025em b}\kern-.08em
    T\kern-.1667em\lower.7ex\hbox{E}\kern-.125emX}}
\begin{document}

\title{\LARGE{PiDRAM: An FPGA-based Framework\\for End-to-end Evaluation of Processing-in-DRAM Techniques}\vspace{-0mm}}

\author{
{Ataberk Olgun\affilETH}\qquad%
{Juan Gomez Luna\affilETH}\qquad
{Konstantinos Kanellopoulos\affilETH}\qquad
{Behzad Salami\affilETH}\qquad \\%
{Hasan Hassan\affilETH}\qquad%
{O\u{g}uz Ergin\affilETU}\qquad
{Onur Mutlu\affilETH}\qquad\vspace{-3mm}\\\\
{\vspace{-3mm}\affilETH \emph{ETH Z{\"u}rich}} \qquad \affilETU \emph{TOBB University of Economics and Technology} \qquad
\vspace{5mm}
}

\maketitle

\fancyhead{}
\thispagestyle{plain}
\pagestyle{plain}

\section{Background}

DRAM-based main memory is used in nearly all computing systems as a major component. Modern memory\rev{-}intensive workloads have increasing memory bandwidth, latency, and capacity requirements. However, DRAM vendors \rev{often} prioritize memory capacity scaling over latency and bandwidth~\cite{chang.sigmetrics16,lee.hpca13,mutlu2020modern,patel2022case}. As a result, main memory \rev{is} an increasingly worsening bottleneck in computing systems~\cite{mutlu2020modern,ghose2019processing,mutlu.imw13,mutlu.superfri15,mutlu2015main,mutlu2019}.

One way of overcoming the main memory bottleneck is to move computation near memory, {a paradigm known as \emph{processing-in-memory} (PiM)}~\cite{fernandez2020natsa,cali2020genasm,kim.bmc18,ahn.pei.isca15,ahn.tesseract.isca15,boroumand.asplos18,boroumand2019conda,boroumand2016pim,singh2019napel,asghari-moghaddam.micro16,JAFAR,farmahini2015nda,gao.pact15,DBLP:conf/hpca/GaoK16,gu.isca16,hashemi.isca16,cont-runahead,hsieh.isca16,kim.isca16,kim.sc17,liu-spaa17,morad.taco15,nai2017graphpim,pattnaik.pact16,pugsley2014ndc,zhang.hpdc14,zhu2013accelerating,DBLP:conf/isca/AkinFH15,gao2017tetris,drumond2017mondrian,dai2018graphh,zhang2018graphp,huang2020heterogeneous,zhuo2019graphq,syncron,aga.hpca17,eckert2018neural,fujiki2019duality,kang.icassp14,chang.hpca16,seshadri.micro17,seshadri2013rowclone,angizi2019graphide,li.dac16,angizi2018pima,angizi2018cmp,angizi2019dna,levy.microelec14,kvatinsky.tcasii14,shafiee2016isaac,kvatinsky.iccd11,kvatinsky.tvlsi14,gaillardon2016plim,bhattacharjee2017revamp,hamdioui2015memristor,xie2015fast,hamdioui2017myth,yu2018memristive,rezaei2020nom,wang2020figaro,mandelman.ibmjrd02,xin2020elp2im,gao2020computedram,li.micro17,deng.dac2018,kim.hpca18,kim.hpca19,hajinazarsimdram,ali2019memory,ronen2022bitlet,zha2019liquid,testa2016inversion,borghetti2010memristive,intel-loihi,geoffrey2017neuromorphic,seshadri.thesis16,Seshadri:2015:ANDOR,seshadri.arxiv16,seshadri.bookchapter17,seshadri.bookchapter17.arxiv, guo2014wondp, cho2020mcdram, ferreira2021pluto, kang1999flexram, giannoula2022sigmetrics, murphy2001characterization, landgraf2021combining, ahmed2019compiler, singh2021fpga, rodrigues2016scattergather, ghiasi2022genstore, deoliveira2021IEEE, jain2018computing, boroumand2021google_arxiv, Zois2018, boroumand2021google, Kautz1969, nair2015evolution, lloyd2017keyvalue, santos2017operand, zha2020hyper, gu2020ipim, asgarifafnir, giannoula2022sparsep, boroumand2021icde, xi2020memory, balasubramonian2014near, herruzo2021enabling, jacob2016compiling, fujiki2018memory, gokhale1995processing, elliott1999computational, nair2015active, lloyd2018dse, singh2020nero, lavenier2020, kogge1994, shaw1981non, Mai:2000:SMM:339647.339673, besta2021sisa, patterson1997case, yavits2021giraf, boroumand2021polynesia, impica, stone1970logic, denzler2021casper, mutlu.imw13, lloyd2015memory, sura2015data, wen2017rebooting, chi2016prime, mutlu.superfri15, deoliveira2021, diab2022hicomb, Draper:2002:ADP:514191.514197, singh2021accelerating, shin2018mcdram, gokhale2015rearr, oskin1998active, zheng2016tcam, amiraliphd, DBLP:conf/sigmod/BabarinsaI15, diab2022high}. {PiM} reduces memory latency between the {memory units and the compute units}, enables the {compute units} to {exploit the large internal bandwidth within} memory devices, and reduces the overall power consumption of the system by eliminating the need for transferring data over power-hungry off-chip interfaces~\cite{mutlu2020modern,ghose2019processing,mutlu2019}. 

PiM techniques provide significant performance benefits and energy savings by exploiting the high internal bit-level parallelism of DRAM for different arithmetic~\cite{seshadri.micro17,xin2020elp2im,hajinazarsimdram,deng.dac2018,li.micro17,angizi2019graphide,ali2019memory,kim.hpca19,olgun2021quactrng,kim.hpca18} and data movement operations~\cite{seshadri2013rowclone,chang.hpca16,rezaei2020nom,wang2020figaro}. Recent works~\cite{gao2020computedram,kim.hpca19,kim.hpca18,olgun2021quactrng,talukder2019exploiting} show that {some} of these PiM techniques {can already be} supported in contemporary, {off-the-shelf} DRAM chips \rev{with modifications only to the memory controller}. Given that DRAM is a {dominant memory technology}, \revt{such commodity DRAM based} PiM techniques provide a promising way to improve the performance and energy efficiency of existing and future systems at \emph{no additional {DRAM} hardware cost}. 

\section{Motivation}

{Integration of PiM techniques in a real system imposes non-trivial challenges that require further research to find appropriate solutions.}
{For example, in-DRAM copy and initialization techniques~\cite{seshadri2013rowclone,gao2020computedram} \revz{that copy data in bulk (e.g., one or multiple memory pages)} require modifications \rev{related to} memory management that affect \rev{various} different parts of the system. 
First, \rev{in-DRAM copy and initialization} techniques have specific memory allocation and alignment requirements (e.g., \revz{page-granularity} source and destination operands should be allocated and aligned \rev{at row granularity within the same} DRAM subarray). These requirements are \emph{not} satisfied by existing \rev{system-level} memory allocation primitives (e.g., \texttt{malloc}~\cite{malloc}, \texttt{posix\_memalign}~\cite{posixmemalign}). 
Second, the source operand\revz{s (usually one or more pages)} of a copy operation must be up to date in DRAM (i.e., \rev{it should} not have dirty copies in CPU caches). Existing cache \revt{coherence} management operations (e.g., \rev{the} x86 CLFLUSH instruction~\cite{x86-manual}) cannot efficiently evict \revz{page-granularity} source operands from the caches~\cite{seshadri2014dirty,seshadri.thesis16}\rev{, as cache \revt{coherence} management operations need to query the coherence states of \emph{\revz{all}} cache blocks of the source operands (i.e., even the coherence states of the cache blocks that are not present in CPU caches are queried)}~\cite{seshadri2014dirty}.}

\subsection{Limitations of the State-of-the-Art}

{\revz{S}ystem integration challenges of PiM techniques can be \rev{efficiently} studied in existing \revt{general-purpose} computing systems \rev{(e.g., personal computers, cloud computers, embedded systems)}, special-purpose testing platforms \rev(e.g., SoftMC~\cite{hassan2017softmc}), or system simulators \rev{(e.g., gem5~\cite{gem5-gpu,GEM5}, Ramulator~\cite{ramulator.github,ramulator}, \revt{Ramulator-PIM~\cite{ramulator-pim},} zsim~\cite{zsim})}.} 
{First, many \emph{\revz{commodity DRAM based PiM techniques}}\footnote{\revt{\revd{Commodity DRAM based PiM techniques are} PiM techniques that can already be supported in \revz{existing} off-the-shelf DRAM chips with \revd{hardware} modifications \revd{\emph{only}} to the memory controller.}} rely on non-standard DDRx operation, where manufacturer-recommended timing parameters for DDRx commands} {are violated~\cite{gao2020computedram,kim.hpca18,kim.hpca19,olgun2021quactrng,talukder2019exploiting}} \rev{(or otherwise new DRAM commands are added, which requires new chip designs and interfaces)}.\footnote{\rev{We are especially interested in PiM techniques that do \emph{not} require \revt{any} modification to the DRAM chips or the DRAM interface.}} 
\revt{Existing general-purpose} computing systems do \emph{not} permit dynamically changing DDRx timing parameters~\cite{lee.hpca15,kim2018solar,chang.sigmetrics16,kim.hpca19,hassan2017softmc}, which is required to integrate {these PiM techniques} into real systems. 
Second, prior works~\cite{kim.hpca18,kim.hpca19,gao2020computedram} show that the reliability of \rev{commodity DRAM based} PiM techniques is highly dependent on environmental {conditions} such as temperature and voltage fluctuations. These effects are exacerbated by the non-standard behavior of {PiM} techniques in real DRAM chips. 
Although special purpose testing platforms \rev{(e.g., SoftMC~\cite{hassan2017softmc})} can be used to conduct {reliability studies}, these platforms do \emph{not} model {an end-to-end} computing system, {where system integration of PiM techniques can be studied.} 
Third, system simulators \rev{(e.g., gem5~\cite{gem5-gpu,GEM5}, Ramulator~\cite{ramulator.github,ramulator}, \revt{Ramulator-PIM~\cite{ramulator-pim},} zsim~\cite{zsim})} can model {end-to-end computing systems}. However, 
they {(i) do \emph{not} model DRAM operation beyond manufacturer-recommended timing \rev{parameters}, (ii) do \emph{not} have a way of interfacing with real DRAM chips that \revz{embody undisclosed and unique} characteristics \revz{that ha\revd{ve} implications on how PiM techniques are integrated into real systems} (e.g., \revd{proprietary and chip-specific} DRAM internal address mapping~\cite{cojocar2020susceptible,salp}), and (iii) \emph{cannot} support studies} on the reliability {of} {PiM} techniques {since system simulators do \emph{not} model environmental conditions.}

\begin{figure*}[!b]
    \vspace{3mm}
    \centering
    \includegraphics[width=0.9\textwidth]{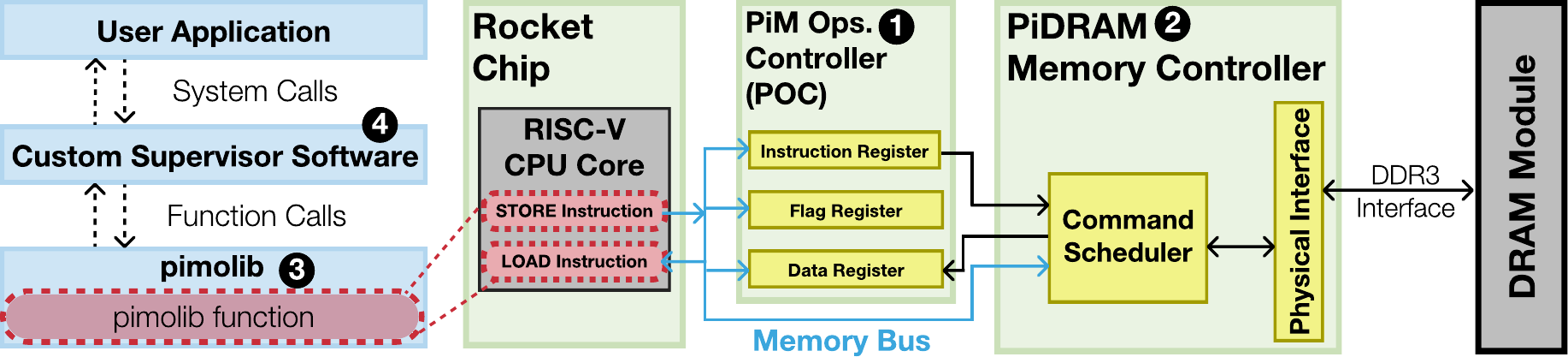}
    \caption{PiDRAM overview. \revd{Modified h}ardware \revt{(in green)} and software \revt{(in blue)} components. \revd{Unmodified components are in gray.} \revt{A pimolib function executes load and store instructions in the CPU to perform PiM operations (in red).} \revt{We use yellow to highlight the key hardware structures that are controlled by the user to perform PiM operations.}}
    \label{fig:overview}
\end{figure*}

\section{PiDRAM Framework}

{Our goal is to design and implement a flexible framework that can be used to solve system integration challenges and analyze trade-offs of end-to-end implementations of \rev{commodity DRAM based} PiM techniques. To this end, we develop \revd{the PiDRAM (\textbf{P}rocessing-\textbf{i}n-\textbf{DRAM})} framework, {the first flexible, end-to-end, and open source framework that enables system integration studies and evaluation of real PiM techniques using real \rev{unmodified} DRAM chips.} 

PiDRAM facilitates system integration studies of \rev{commodity DRAM based} PiM techniques by providing a set of {four} customizable hardware and software components that can be used as {a} common basis to enable system support for such techniques in real systems. Figure~\ref{fig:overview} presents an overview of PiDRAM's components.}

\subsection{Hardware Components}

PiDRAM comprises two key hardware components. Both of these components are designed with the goal to provide a flexible and easy to use framework for evaluating PiM techniques.

\noindent
\textbf{\ding{182} PiM Operations Controller (POC).} POC decodes and executes PiDRAM instructions (e.g., RowClone-Copy~\cite{seshadri2013rowclone}) that are used by the programmer to perform PiM operations. POC communicates with the rest of the system over two well\rev{-}defined interfaces. First, it communicates with the CPU over a memory-mapped interface\rev{, where the CPU can send data to or receive data from POC using memory store and load instructions}. The CPU accesses the memory-mapped registers (\emph{instruction}, \emph{data}, and \emph{flag} registers) in POC to execute in-DRAM operations. This way, we improve the portability of the framework and facilitate porting it to systems that employ different instruction set architectures. Second, \rev{POC} communicates with the memory controller to perform PiM operations in the DRAM chip over a simple hardware interface. \rev{To do so,} POC (i) requests the memory controller to perform a PiM operation, (ii) waits until the memory controller performs the operation, and (iii) receives the result of the PiM operation from the memory controller. \rev{The CPU can read the result of the operation by executing load instructions that target the \emph{data} register in POC.}

\noindent
\textbf{\ding{183} PiDRAM Memory Controller.} PiDRAM's memory controller provides an easy\revt{-}to\revt{-}extend basis for \revt{commodity DRAM based} PiM techniques that require issuing DRAM commands with violated timing parameters~\cite{gao2020computedram,kim.hpca19,kim.hpca18,olgun2021quactrng,talukder2019exploiting}. 
The memory controller is designed modularly and requires \rev{easy\revt{-}to\revt{-}make} modifications to its scheduler to implement new PiM techniques. For instance, our modular design enables supporting RowClone~\cite{seshadri2013rowclone} operations \revt{using} \rev{only} 60 \revt{additional} lines of Verilog code on top of the baseline custom memory controller's scheduler that implements conventional DRAM operations~\cite{salp,lee.hpca13,donghyuk-ddma,chang.sigmetrics17,ghose2018vampire,patel2017reaper,luo2020clr,ghose2019demystifying,kevinchang-thesis,yoongu-thesis,lee.thesis16} (e.g., \revt{activate, precharge,} read, write).

\subsection{Software Components}

PiDRAM comprises two key software components that complement \rev{and control} PiDRAM's hardware components \rev{to} provid\rev{e} a flexible and easy to use \rev{end-to-end} PiM framework. 

\noindent
\textbf{\ding{184} Extensible Software Library \revt{(pimolib)}.} The extensible library (\revt{\underline{PiM} \underline{o}perations \underline{lib}rary}) allows system designers to implement software support for PiM techniques. \revt{Pimolib} contains customizable functions that interface with the POC to perform PiM operations in real \rev{unmodified} DRAM chips. A typical function in \revt{pimolib} performs a PiM operation in four steps: It (i) writes a PiDRAM instruction to \rev{the} POC's \emph{instruction} register, (ii) sets the \emph{Start} \revt{flag} in the POC's \emph{flag} register, (iii) waits for \rev{the} POC to set the \emph{Ack} \revt{flag} in the POC's \emph{flag} register, and (iv) reads the result of the PiM operation from the POC's \emph{data} register \revt{(e.g., the true random number after performing a\revz{n in-DRAM true random number generation} operation, Section~\ref{sec:case-studies})}.

\noindent
\textbf{\ding{185} Custom Supervisor Software.} The supervisor software \rev{contains} the necessary OS primitives (e.g., memory management, allocation, and alignment \rev{specialized for RowClone~\cite{seshadri2013rowclone}}). This facilitates developing end-to-end integration of PiM techniques \rev{in the system} as these techniques require modifications across the software stack. \rev{For example, integrating RowClone end-to-end \revt{in the full system} requires a new memory allocation mechanism (Section~\ref{sec:case-studies}) that can satisfy the memory allocation constraints of RowClone~\cite{seshadri2013rowclone,pidramarxiv}.}

\begin{figure*}[!t]
    \centering
    \includegraphics[width=0.9\textwidth]{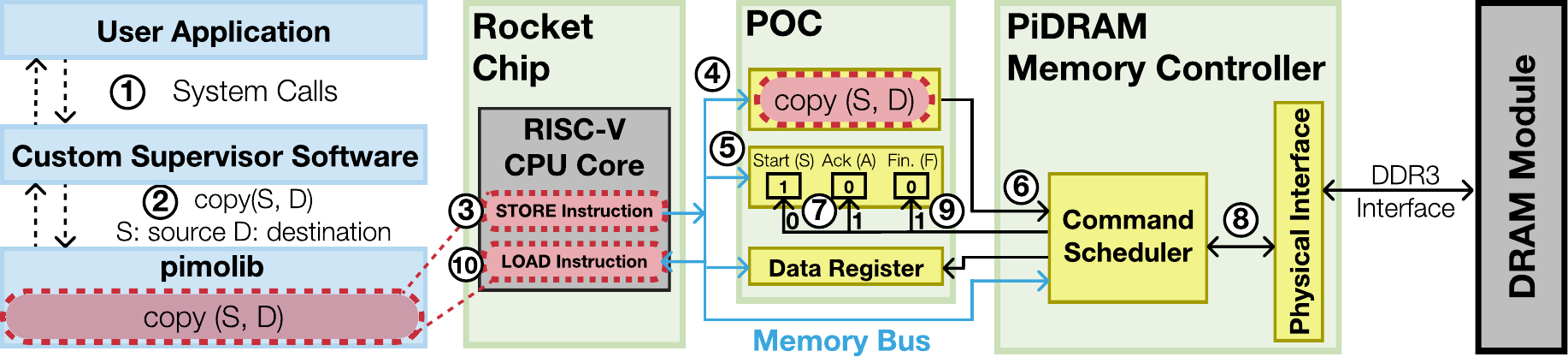}
    \caption{\revt{Workflow for a PiDRAM RowClone-Copy operation.}}
    \label{fig:flow}
\end{figure*}

\subsection{General PiDRAM Execution Workflow}

\rev{We describe the general workflow for a PiDRAM operation (e.g., RowClone-Copy~\cite{seshadri2013rowclone}, random number generation using D-RaNGe~\cite{kim.hpca19}) in Figure~\ref{fig:flow} over an example \texttt{\revt{copy()}} function that is called by the user to \revt{perform a RowClone-Copy operation} in DRAM.}

\revt{T}he user makes a system call to the custom supervisor software \revt{\ding{172}} that in turn calls the \revt{\texttt{copy(source, destination)}} function in the \revz{pimolib} \revt{\ding{173}}. The function executes \revd{two} store instructions in the RISC-V core \ding{174}. \revd{The first store} instruction \revt{update\revd{s}} the \emph{instruction} register with the \revt{copy} instruction (i.e., the instruction that performs a \revt{RowClone-Copy} operation in DRAM) \revt{\ding{175}} and \revd{the second store instruction} \revt{set\revd{s} the Start flag in the flag register to logic-1 \ding{176} in the POC.} \revt{When the Start flag is set,} the POC instructs the PiDRAM memory controller to perform a \revt{RowClone-Copy} operation using violated timing parameters \ding{177}. \revt{\revz{T}he \revz{POC waits until the memory controller starts executing the \revd{operation, after which it}} sets the Start flag to logic-0 and the Ack \revz{flag} to logic-1 \ding{178}\revd{, indicating that it started the execution of the PiM operation}.} \revd{T}he PiDRAM memory controller performs the \revt{RowClone-Copy} operation by issuing a set of DRAM commands with violated timing parameters \revt{\ding{179}}. \revt{\revz{When the last DRAM command is issued, the memory controller} sets the Finish flag (denoted as Fin. in Figure~\ref{fig:flow}) in the flag register to logic-1 \revd{\ding{180}}, indicating the end of execution for the last PiM operation that the memory controller acknowledged.} \revt{The copy function periodically checks \revz{either} the Ack \revz{or the} Finish flag in the flag register \revz{(depending on a user-supplied argument)} by executing load instructions that target the flag register \ding{181}. \revz{When the periodically checked flag is set, the copy function returns.} This way, the copy function optionally blocks until the start \revz{(i.e., the Ack flag is set)} or the end \revz{(i.e., the Finish flag is set)} of the execution of the PiM operation (in this example, RowClone-Copy).\footnote{\revz{The data register is not used in RowClone-Copy~\cite{seshadri2013rowclone} operations because the result of the RowClone-Copy operation is stored \revd{\emph{in memory}} (i.e., the source \revd{memory row} is copied to the destination \revd{memory row}). The data register is used in D-RaNGe~\cite{kim.hpca19} operations\revd{, as described in~\cite{pidramarxiv}}. \revd{When used, t}he command scheduler store\revd{s} the random numbers generated by \revd{the} D-RaNGe operation in the data register. To read the generated random numbers, we implement a pimolib function \revd{called} \texttt{rand\_dram()} that executes load instructions in the \revd{RISC-V} core to retrieve the random numbers from the data register in the POC.}}}

\section{FPGA Prototype \& Case Studies}
\label{sec:prototype}
\rev{PiDRAM allows (i) interfacing with and performing in-DRAM operations on real \revt{unmodified} DRAM chips, (ii) observing the effects of environmental conditions on PiM techniques, and (iii) conducting end-to-end \revt{system-level} studies on PiM techniques using real DRAM chips. We demonstrate the versatility and ease-of-use of PiDRAM by prototyping it on an FPGA board and conducting two case studies \revt{of PIM computation}.}

\subsection{\rev{FPGA Prototype}}
\begin{figure}[h]
    \centering
    \includegraphics[width=0.48\textwidth]{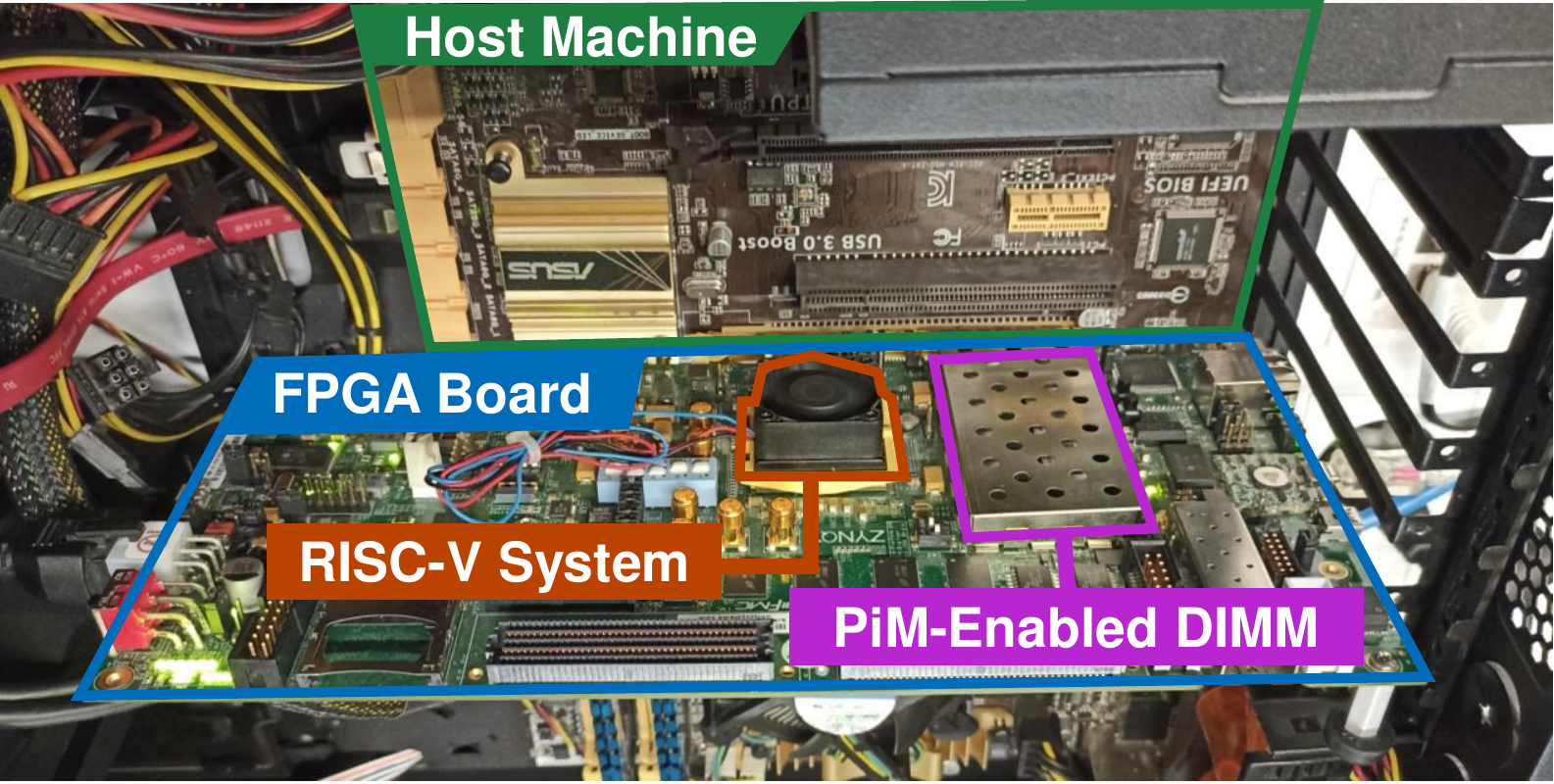}
    \caption{PiDRAM's \revv{FPGA} prototype.}
    \label{fig:picture}
    \vspace{-3mm}
\end{figure}

We demonstrate a prototype of PiDRAM on an FPGA-based platform (Xilinx ZC706~\cite{zc706}) that implements an open-source RISC-V system (Rocket Chip~\cite{asanovic2016rocket}). Our custom supervisor software extends the RISC-V PK~\cite{riscv-pk} to support the necessary OS primitives on PiDRAM's prototype. We perform in-DRAM operations in real \rev{unmodified} DDR3 DRAM chips that are connected to the FPGA-based platform. \revd{Figure~\ref{fig:picture} shows our FPGA prototype.}

\subsection{\rev{Case Studies}}
\label{sec:case-studies}
To demonstrate the flexibility and ease of use of PiDRAM, we implement two PiM {techniques} \rev{on unmodified DRAM chips by violating timing parameters}: (1) \emph{RowClone}~\cite{seshadri2013rowclone}, {an in-DRAM copy and initialization mechanism (using command sequences proposed by ComputeDRAM~\cite{gao2020computedram})}, and (2) \emph{D-RaNGe}~\cite{kim.hpca19}, {an in-DRAM true random number generator based on DRAM activation-latency failure{s}} \revt{induced by violation of manufacturer-recommended DRAM timing parameters}. 

{\revt{T}o support RowClone}, we (i) {customize} PiDRAM's memory controller {to} issue carefully-engineered sequences of DRAM commands that perform copy operations in DRAM, {and} (ii) extend the custom supervisor software to implement a new memory management mechanism {that} satisfies the memory allocation {and alignment} requirements of RowClone~\cite{pidramarxiv,seshadri2013rowclone}. The key idea of the memory management mechanism is to allocate physical addresses in the same DRAM subarray~\cite{salp,seshadri2013rowclone,seshadri.micro17,chang.hpca16} for the source and destination operands \rev{(i.e., rows)} of the copy operation. We develop a new methodology \rev{that} check\rev{s} if RowClone operations on randomly\rev{-}generated physical addresses succeed or fail, to find physical \rev{row} addresses that \rev{are in} the same DRAM subarray. \rev{We explain this methodology \revt{and our techniques} in more detail in our arXiv paper~\cite{pidramarxiv}.}

To support D-RaNGe, we (i) make simple modifications to \rev{the} PiDRAM memory controller to implement a new storage structure that buffers random numbers (\revt{i.e., a} random number buffer) and (ii) extend the software library with functions that retrieve data from the random number buffer \revt{and supply it to} the user program. 
 
\section{Key Results \& Contributions}

Our evaluation of RowClone \rev{shows that an end-to-end implementation of RowClone achieves (i) {118.5$\times{}$} speedup for copy and {88.7$\times{}$} \revt{speedup for} initialization operations\revt{,} assuming data in DRAM is up to date and the source operand is not cached in CPU caches (i.e., no cache \revt{coherence} management operations are required)} and (ii) {14.6$\times{}$} speedup for copy and {12.6$\times{}$} initialization operations \rev{assuming data in DRAM is up to date but the source operand has cached copies in CPU caches (i.e., a cache \revt{coherence} management operation must be performed for every cache block of the source operand)} over CPU\revt{-based} copy (i.e., conventional \texttt{memcpy}~\cite{memcpy}) and initialization (i.e., conventional \texttt{calloc}~\cite{calloc}) operations. 

\revt{We make two observations from o}ur \rev{evaluation} of D-RaNGe. \revt{First, end-to-end integration of D-RaNGe \revz{into a real system}} provides true random numbers at low latency. \revt{Our implementation can generate a 4-bit random number in \SI{220}{\nano\second}.} \revt{Second, an end-to-end integration of D-RaNGe \revz{into a real system} generates true random numbers at high throughput.} \revt{Our implementation achieves} {8.30} Mb/s \rev{sustained throughput}. \revt{PiDRAM's D-RaNGe implementation can be optimized to generate random numbers more frequently (i.e., at higher throughput). We leave such optimizations to PiDRAM's D-RaNGe implementation for future work.}

Over the Verilog and C++ \rev{codebase} provided by PiDRAM, \rev{integrating} RowClone end-to-end \revt{in the full system} takes 198 \revt{lines of Verilog and 565 lines of C++ code} and \revt{integrating D-RaNGe end-to-end in the full system takes 190 lines of Verilog and 78 lines of C++ code}.


Our contributions are as follows:

\begin{itemize}
  \item We develop PiDRAM, the first flexible framework that enables end-to-end integration and evaluation of PiM {techniques} using real \rev{unmodified} DRAM chips.
  
  \item We develop a prototype of PiDRAM on an FPGA-based platform. To demonstrate the ease-of-use \revt{and evaluation benefits} of PiDRAM, we implement two state-of-the-art DRAM-based PiM {techniques, RowClone and D-RaNGe,} and evaluate them on PiDRAM's prototype \rev{using unmodified DDR3 DRAM chips}. 
  
  \item We devise a new memory management mechanism that satisfies the memory allocation {and alignment} requirements of RowClone~\cite{seshadri2013rowclone}. We demonstrate that our mechanism enables RowClone end-to-end \revt{in the full system}, \rev{and} {provid\rev{es}} significant performance improvements {over traditional CPU\revt{-based} copy \revt{and initialization} \rev{operations \revt{(\texttt{memcpy}~\cite{memcpy} and \texttt{calloc}~\cite{calloc})} as demonstrated on our PiDRAM prototype.}}
  
  \item {We implement and evaluate a \rev{state-of-the-art} DRAM true random number generat\revt{ion technique} (D-RaNGe~\cite{kim.hpca19}). Our implementation provides a \revt{solid foundation} for future work on system integration of DRAM-based PiM security primitives (e.g., PUFs~\cite{talukder2019exploiting,kim.hpca18}, TRNG\revz{s\cite{kim.hpca19,olgun2021quactrng,talukder2019exploiting}}), \rev{implemented using real unmodified DRAM chips}.}
  
  \item{\rev{PiDRAM is open sourced \rev{and freely available} on Github: \url{https://github.com/CMU-SAFARI/PiDRAM}.}}
\end{itemize}

\section{\rev{Summary}}
We introduce the first flexible, end-to-end, and open source in-DRAM computation framework, \rev{called} PiDRAM. \revt{Unlike} existing \revt{evaluation frameworks} and prior work, PiDRAM allows \revt{us to} conduct end-to-end \revt{full system} studies o\revt{f} \rev{commodity DRAM based} PiM techniques \revt{that use real unmodified DRAM chips}. We demonstrate PiDRAM's versatility and ease of use by prototyp\revt{ing it} on an FPGA board and conducting case studies o\revz{f} \rev{two state-of-the-art} PiM techniques \revz{for} \revt{in-DRAM data copy \& initialization (RowClone~\cite{seshadri2013rowclone}) and in-DRAM true random number generation (D-RaNGe~\cite{kim.hpca19})}\revz{.} We show that an end-to-end implementation of (i) RowClone greatly improve\revt{s} \rev{data} copy \revt{and} initialization \revt{performance at the full system level}, and (ii) D-RaNGe provide\revt{s} true random numbers to applications at high throughput. \rev{We hope and believe that researchers and industry will benefit from and build on PiDRAM \revt{to better evaluate PiM techniques and their benefits in real systems}.}  

\section*{Acknowledgements}

We thank the SAFARI Research Group members for valuable feedback and the stimulating intellectual environment they provide. We acknowledge the generous gifts provided by our industrial partners{, including} Google, Huawei, Intel, Microsoft, and VMware. \rev{We acknowledge support from the Semiconductor Research Corporation and the ETH Future Computing Laboratory.} 

This invited extended abstract is a summary version of our prior work~\cite{pidramarxiv}. A presentation that describes \rev{this} work can be found at~\cite{youtube}.

\balance 
\bibliographystyle{IEEEtran}
\bibliography{PiDRAM}

\begin{thebibliography}{100}
\providecommand{\url}[1]{#1}
\csname url@samestyle\endcsname
\providecommand{\newblock}{\relax}
\providecommand{\bibinfo}[2]{#2}
\providecommand{\BIBentrySTDinterwordspacing}{\spaceskip=0pt\relax}
\providecommand{\BIBentryALTinterwordstretchfactor}{4}
\providecommand{\BIBentryALTinterwordspacing}{\spaceskip=\fontdimen2\font plus
\BIBentryALTinterwordstretchfactor\fontdimen3\font minus
  \fontdimen4\font\relax}
\providecommand{\BIBforeignlanguage}[2]{{%
\expandafter\ifx\csname l@#1\endcsname\relax
\typeout{** WARNING: IEEEtran.bst: No hyphenation pattern has been}%
\typeout{** loaded for the language `#1'. Using the pattern for}%
\typeout{** the default language instead.}%
\else
\language=\csname l@#1\endcsname
\fi
#2}}
\providecommand{\BIBdecl}{\relax}
\BIBdecl

\bibitem{chang.sigmetrics16}
K.~K. Chang \emph{et~al.}, ``{Understanding Latency Variation in Modern DRAM
  Chips: Experimental Characterization, Analysis, and Optimization},'' in
  \emph{SIGMETRICS}, 2016.

\bibitem{lee.hpca13}
D.~Lee \emph{et~al.}, ``{Tiered-Latency DRAM: A Low Latency and Low Cost DRAM
  Architecture},'' in \emph{HPCA}, 2013.

\bibitem{mutlu2020modern}
O.~Mutlu \emph{et~al.}, ``{A Modern Primer on Processing in Memory},''
  \emph{Emerging Computing: From Devices to Systems - Looking Beyond Moore and
  Von Neumann}, 2021.

\bibitem{patel2022case}
M.~Patel \emph{et~al.}, ``{A Case for Transparent Reliability in DRAM
  Systems},'' arXiv:2204.10378, 2022.

\bibitem{ghose2019processing}
S.~Ghose \emph{et~al.}, ``{Processing-in-Memory: A Workload-driven
  Perspective},'' \emph{IBM JRD}, 2019.

\bibitem{mutlu.imw13}
O.~Mutlu, ``{Memory Scaling: A Systems Architecture Perspective},'' \emph{IMW},
  2013.

\bibitem{mutlu.superfri15}
O.~Mutlu \emph{et~al.}, ``{Research Problems and Opportunities in Memory
  Systems},'' \emph{SUPERFRI}, 2014.

\bibitem{mutlu2015main}
O.~Mutlu, ``{Main Memory Scaling: Challenges and Solution Directions},'' 2015,
  {Invited Book Chapter in More than Moore Technologies for Next Generation
  Computer Design, pp. 127-153, Springer}.

\bibitem{mutlu2019}
O.~Mutlu \emph{et~al.}, ``{Processing Data Where It Makes Sense: {E}nabling
  In-Memory Computation},'' \emph{MicPro}, 2019.

\bibitem{fernandez2020natsa}
I.~Fernandez \emph{et~al.}, ``{NATSA: A Near-Data Processing Accelerator for
  Time Series Analysis},'' in \emph{ICCD}, 2020.

\bibitem{cali2020genasm}
D.~S. Cali \emph{et~al.}, ``{GenASM: A High-Performance, Low-Power Approximate
  String Matching Acceleration Framework for Genome Sequence Analysis},'' in
  \emph{MICRO}, 2020.

\bibitem{kim.bmc18}
J.~S. Kim \emph{et~al.}, ``{GRIM-Filter: Fast Seed Location Filtering in DNA
  Read Mapping Using Processing-in-Memory Technologies},'' \emph{BMC Genomics},
  2018.

\bibitem{ahn.pei.isca15}
J.~Ahn \emph{et~al.}, ``{PIM-Enabled Instructions: A Low-Overhead,
  Locality-Aware Processing-in-Memory Architecture},'' in \emph{ISCA}, 2015.

\bibitem{ahn.tesseract.isca15}
J.~Ahn \emph{et~al.}, ``{A Scalable Processing-in-Memory Accelerator for
  Parallel Graph Processing},'' in \emph{ISCA}, 2015.

\bibitem{boroumand.asplos18}
A.~Boroumand \emph{et~al.}, ``{Google Workloads for Consumer Devices:
  Mitigating Data Movement Bottlenecks},'' in \emph{ASPLOS}, 2018.

\bibitem{boroumand2019conda}
A.~Boroumand \emph{et~al.}, ``{CoNDA: Efficient Cache Coherence Support for
  near-Data Accelerators},'' in \emph{ISCA}, 2019.

\bibitem{boroumand2016pim}
A.~Boroumand \emph{et~al.}, ``{LazyPIM: An Efficient Cache Coherence Mechanism
  for Processing-in-Memory},'' \emph{CAL}, 2016.

\bibitem{singh2019napel}
G.~Singh \emph{et~al.}, ``{NAPEL: Near-memory Computing Application Performance
  Prediction via Ensemble Learning},'' in \emph{DAC}, 2019.

\bibitem{asghari-moghaddam.micro16}
H.~Asghari-Moghaddam \emph{et~al.}, ``{Chameleon: Versatile and Practical
  Near-DRAM Acceleration Architecture for Large Memory Systems},'' in
  \emph{MICRO}, 2016.

\bibitem{JAFAR}
S.~L. Xi \emph{et~al.}, ``{Beyond the Wall: Near-Data Processing for
  Databases},'' in \emph{DaMoN}, 2015.

\bibitem{farmahini2015nda}
A.~Farmahini-Farahani \emph{et~al.}, ``{NDA: Near-DRAM acceleration
  architecture leveraging commodity DRAM devices and standard memory
  modules},'' in \emph{HPCA}, 2015.

\bibitem{gao.pact15}
M.~Gao \emph{et~al.}, ``{Practical Near-Data Processing for In-Memory Analytics
  Frameworks},'' in \emph{PACT}, 2015.

\bibitem{DBLP:conf/hpca/GaoK16}
M.~Gao \emph{et~al.}, ``{HRL: Efficient and Flexible Reconfigurable Logic for
  Near-Data Processing},'' in \emph{HPCA}, 2016.

\bibitem{gu.isca16}
B.~Gu \emph{et~al.}, ``{Biscuit: {A} Framework for Near-Data Processing of Big
  Data Workloads},'' in \emph{ISCA}, 2016.

\bibitem{hashemi.isca16}
M.~Hashemi \emph{et~al.}, ``{Accelerating Dependent Cache Misses with an
  Enhanced Memory Controller},'' in \emph{ISCA}, 2016.

\bibitem{cont-runahead}
M.~Hashemi \emph{et~al.}, ``{Continuous Runahead: Transparent Hardware
  Acceleration for Memory Intensive Workloads},'' in \emph{MICRO}, 2016.

\bibitem{hsieh.isca16}
K.~Hsieh \emph{et~al.}, ``{Transparent Offloading and Mapping (TOM): Enabling
  Programmer-Transparent Near-Data Processing in GPU Systems},'' in
  \emph{ISCA}, 2016.

\bibitem{kim.isca16}
D.~Kim \emph{et~al.}, ``{Neurocube: {A} Programmable Digital Neuromorphic
  Architecture with High-Density {3D} Memory},'' in \emph{ISCA}, 2016.

\bibitem{kim.sc17}
G.~Kim \emph{et~al.}, ``{Toward Standardized Near-Data Processing with
  Unrestricted Data Placement for GPUs},'' in \emph{SC}, 2017.

\bibitem{liu-spaa17}
Z.~Liu \emph{et~al.}, ``{Concurrent Data Structures for Near-Memory
  Computing},'' in \emph{SPAA}, 2017.

\bibitem{morad.taco15}
A.~Morad \emph{et~al.}, ``{GP-SIMD Processing-in-Memory},'' \emph{ACM TACO},
  2015.

\bibitem{nai2017graphpim}
L.~Nai \emph{et~al.}, ``{GraphPIM: Enabling Instruction-Level PIM Offloading in
  Graph Computing Frameworks},'' in \emph{HPCA}, 2017.

\bibitem{pattnaik.pact16}
A.~Pattnaik \emph{et~al.}, ``{Scheduling Techniques for GPU Architectures with
  Processing-in-Memory Capabilities},'' in \emph{PACT}, 2016.

\bibitem{pugsley2014ndc}
S.~H. Pugsley \emph{et~al.}, ``{{NDC: Analyzing the Impact of 3D-Stacked
  Memory+Logic Devices on MapReduce Workloads}},'' in \emph{ISPASS}, 2014.

\bibitem{zhang.hpdc14}
D.~P. Zhang \emph{et~al.}, ``{TOP-PIM: Throughput-Oriented Programmable
  Processing in Memory},'' in \emph{HPDC}, 2014.

\bibitem{zhu2013accelerating}
Q.~Zhu \emph{et~al.}, ``{Accelerating Sparse Matrix-Matrix Multiplication with
  3D-Stacked Logic-in-Memory Hardware},'' in \emph{HPEC}, 2013.

\bibitem{DBLP:conf/isca/AkinFH15}
B.~Akin \emph{et~al.}, ``{Data Reorganization in Memory Using {3D}-Stacked
  {DRAM}},'' in \emph{ISCA}, 2015.

\bibitem{gao2017tetris}
M.~Gao \emph{et~al.}, ``{Tetris: Scalable and Efficient Neural Network
  Acceleration with 3D Memory},'' in \emph{ASPLOS}, 2017.

\bibitem{drumond2017mondrian}
M.~P. Drumond Lages~de Oliveira \emph{et~al.}, ``{The Mondrian Data Engine},''
  in \emph{ISCA}, 2017.

\bibitem{dai2018graphh}
G.~Dai \emph{et~al.}, ``{GraphH: A Processing-in-Memory Architecture for
  Large-scale Graph Processing},'' \emph{IEEE TCAD}, 2018.

\bibitem{zhang2018graphp}
M.~Zhang \emph{et~al.}, ``{GraphP: Reducing Communication for PIM-based Graph
  Processing with Efficient Data Partition},'' in \emph{HPCA}, 2018.

\bibitem{huang2020heterogeneous}
Y.~Huang \emph{et~al.}, ``{A Heterogeneous PIM Hardware-Software Co-Design for
  Energy-Efficient Graph Processing},'' in \emph{IPDPS}, 2020.

\bibitem{zhuo2019graphq}
Y.~Zhuo \emph{et~al.}, ``{GraphQ: Scalable PIM-based Graph Processing},'' in
  \emph{MICRO}, 2019.

\bibitem{syncron}
C.~Giannoula \emph{et~al.}, ``{SynCron: Efficient Synchronization Support for
  Near-Data-Processing Architectures},'' in \emph{HPCA}, 2021.

\bibitem{aga.hpca17}
S.~Aga \emph{et~al.}, ``{Compute Caches},'' in \emph{HPCA}, 2017.

\bibitem{eckert2018neural}
C.~Eckert \emph{et~al.}, ``{Neural Cache: Bit-serial In-cache Acceleration of
  Deep Neural Networks},'' in \emph{ISCA}, 2018.

\bibitem{fujiki2019duality}
D.~Fujiki \emph{et~al.}, ``{Duality Cache for Data Parallel Acceleration},'' in
  \emph{ISCA}, 2019.

\bibitem{kang.icassp14}
M.~Kang \emph{et~al.}, ``{An Energy-Efficient VLSI Architecture for Pattern
  Recognition via Deep Embedding of Computation in SRAM},'' in \emph{ICASSP},
  2014.

\bibitem{chang.hpca16}
K.~K. Chang \emph{et~al.}, ``{Low-Cost Inter-Linked Subarrays (LISA): Enabling
  Fast Inter-Subarray Data Movement in DRAM},'' in \emph{HPCA}, 2016.

\bibitem{seshadri.micro17}
V.~Seshadri \emph{et~al.}, ``{Ambit: In-Memory Accelerator for Bulk Bitwise
  Operations Using Commodity DRAM Technology},'' in \emph{MICRO}, 2017.

\bibitem{seshadri2013rowclone}
V.~Seshadri \emph{et~al.}, ``{RowClone: Fast and Energy-Efficient In-DRAM Bulk
  Data Copy and Initialization},'' in \emph{MICRO}, 2013.

\bibitem{angizi2019graphide}
S.~Angizi \emph{et~al.}, ``{Graphide: A Graph Processing Accelerator Leveraging
  In-dram-computing},'' in \emph{GLSVLSI}, 2019.

\bibitem{li.dac16}
S.~Li \emph{et~al.}, ``{Pinatubo: A Processing-in-Memory Architecture for Bulk
  Bitwise Operations in Emerging Non-Volatile Memories},'' in \emph{DAC}, 2016.

\bibitem{angizi2018pima}
S.~Angizi \emph{et~al.}, ``{PIMA-Logic: A Novel Processing-in-Memory
  Architecture for Highly Flexible and Energy-efficient Logic Computation},''
  in \emph{DAC}, 2018.

\bibitem{angizi2018cmp}
S.~Angizi \emph{et~al.}, ``{CMP-PIM: An Energy-Efficient Comparator-Based
  Processing-in-Memory Neural Network Accelerator},'' in \emph{DAC}, 2018.

\bibitem{angizi2019dna}
S.~Angizi \emph{et~al.}, ``{AlignS: A Processing-in-Memory Accelerator for DNA
  Short Read Alignment Leveraging SOT-MRAM},'' in \emph{DAC}, 2019.

\bibitem{levy.microelec14}
Y.~Levy \emph{et~al.}, ``{Logic Operations in Memory Using a Memristive Akers
  Array},'' \emph{Microelectronics Journal}, 2014.

\bibitem{kvatinsky.tcasii14}
S.~Kvatinsky \emph{et~al.}, ``{MAGIC---Memristor-Aided Logic},'' \emph{IEEE
  TCAS II: Express Briefs}, 2014.

\bibitem{shafiee2016isaac}
A.~Shafiee \emph{et~al.}, ``{ISAAC: A Convolutional Neural Network Accelerator
  with In-Situ Analog Arithmetic in Crossbars},'' in \emph{ISCA}, 2016.

\bibitem{kvatinsky.iccd11}
S.~Kvatinsky \emph{et~al.}, ``{Memristor-Based IMPLY Logic Design Procedure},''
  in \emph{ICCD}, 2011.

\bibitem{kvatinsky.tvlsi14}
S.~Kvatinsky \emph{et~al.}, ``{Memristor-Based Material Implication (IMPLY)
  Logic: Design Principles and Methodologies},'' \emph{TVLSI}, 2014.

\bibitem{gaillardon2016plim}
P.-E. Gaillardon \emph{et~al.}, ``{The Programmable Logic-in-Memory (PLiM)
  Computer},'' in \emph{DATE}, 2016.

\bibitem{bhattacharjee2017revamp}
D.~Bhattacharjee \emph{et~al.}, ``{ReVAMP: ReRAM based VLIW Architecture for
  In-Memory Computing},'' in \emph{DATE}, 2017.

\bibitem{hamdioui2015memristor}
S.~Hamdioui \emph{et~al.}, ``{Memristor Based Computation-in-Memory
  Architecture for Data-intensive Applications},'' in \emph{DATE}, 2015.

\bibitem{xie2015fast}
L.~Xie \emph{et~al.}, ``{Fast Boolean Logic Mapped on Memristor Crossbar},'' in
  \emph{ICCD}, 2015.

\bibitem{hamdioui2017myth}
S.~Hamdioui \emph{et~al.}, ``{Memristor for Computing: Myth or Reality?}'' in
  \emph{DATE}, 2017.

\bibitem{yu2018memristive}
J.~Yu \emph{et~al.}, ``{Memristive Devices for Computation-in-Memory},'' in
  \emph{DATE}, 2018.

\bibitem{rezaei2020nom}
S.~H.~S. {Rezaei} \emph{et~al.}, ``{NoM: Network-on-Memory for Inter-Bank Data
  Transfer in Highly-Banked Memories},'' \emph{CAL}, 2020.

\bibitem{wang2020figaro}
Y.~Wang \emph{et~al.}, ``{FIGARO: Improving System Performance via Fine-Grained
  In-DRAM Data Relocation and Caching},'' in \emph{MICRO}, 2020.

\bibitem{mandelman.ibmjrd02}
J.~A. Mandelman \emph{et~al.}, ``{Challenges and Future Directions for the
  Scaling of Dynamic Random-Access Memory ({DRAM})},'' \emph{IBM JRD}, 2002.

\bibitem{xin2020elp2im}
X.~Xin \emph{et~al.}, ``{ELP2IM: Efficient and Low Power Bitwise Operation
  Processing in DRAM},'' in \emph{HPCA}, 2020.

\bibitem{gao2020computedram}
F.~Gao \emph{et~al.}, ``{ComputeDRAM: In-Memory Compute Using Off-the-Shelf
  DRAMs},'' in \emph{MICRO}, 2019.

\bibitem{li.micro17}
S.~Li \emph{et~al.}, ``{DRISA: A DRAM-Based Reconfigurable In-Situ
  Accelerator},'' in \emph{MICRO}, 2017.

\bibitem{deng.dac2018}
Q.~Deng \emph{et~al.}, ``{DrAcc: a DRAM Based Accelerator for Accurate CNN
  Inference},'' in \emph{DAC}, 2018.

\bibitem{kim.hpca18}
J.~Kim \emph{et~al.}, ``{The {DRAM} Latency {PUF}: Quickly Evaluating Physical
  Unclonable Functions by Exploiting the Latency--Reliability Tradeoff in
  Modern {DRAM} Devices},'' in \emph{HPCA}, 2018.

\bibitem{kim.hpca19}
J.~Kim \emph{et~al.}, ``{D-RaNGe: Using Commodity {DRAM} Devices to Generate
  True Random Numbers with Low Latency and High Throughput},'' in \emph{HPCA},
  2019.

\bibitem{hajinazarsimdram}
N.~Hajinazar \emph{et~al.}, ``{SIMDRAM: A Framework for Bit-Serial SIMD
  Processing Using DRAM},'' in \emph{ASPLOS}, 2021.

\bibitem{ali2019memory}
M.~F. Ali \emph{et~al.}, ``{In-Memory Low-Cost Bit-Serial Addition Using
  Commodity DRAM Technology},'' in \emph{{TCAS-I}}, 2019.

\bibitem{ronen2022bitlet}
R.~Ronen \emph{et~al.}, ``{The Bitlet Model: A Parameterized Analytical Model
  to Compare PIM and CPU Systems},'' \emph{J. Emerg. Technol. Comput. Syst.},
  2022.

\bibitem{zha2019liquid}
Y.~Zha \emph{et~al.}, ``{Liquid Silicon: A Nonvolatile Fully Programmable
  Processing-In-Memory Processor with Monolithically Integrated ReRAM for Big
  Data/Machine Learning Applications},'' in \emph{Symposium on VLSI Circuits},
  2019.

\bibitem{testa2016inversion}
E.~Testa \emph{et~al.}, ``{Inversion Optimization in Majority-Inverter
  Graphs},'' in \emph{NANOARCH}, 2016.

\bibitem{borghetti2010memristive}
J.~Borghetti \emph{et~al.}, ``{Memristive Switches Enable Stateful Logic
  Operations via Material Implication},'' \emph{Nature}, 2010.

\bibitem{intel-loihi}
Intel, ``{Taking Neuromorphic Computing to the Next Level with Loihi 2},''
  Technology Brief, 2022.

\bibitem{geoffrey2017neuromorphic}
G.~W. Burr \emph{et~al.}, ``{Neuromorphic Computing Using Non-volatile
  Memory},'' \emph{Advances in Physics: X}, 2017.

\bibitem{seshadri.thesis16}
V.~Seshadri, ``{Simple DRAM and Virtual Memory Abstractions to Enable Highly
  Efficient Memory Systems},'' Ph.D. dissertation, Carnegie Mellon University,
  2016.

\bibitem{Seshadri:2015:ANDOR}
V.~Seshadri \emph{et~al.}, ``{Fast Bulk Bitwise AND and OR in DRAM},''
  \emph{CAL}, 2015.

\bibitem{seshadri.arxiv16}
V.~Seshadri \emph{et~al.}, ``{Buddy-RAM: Improving the Performance and
  Efficiency of Bulk Bitwise Operations Using DRAM},'' arXiv:1611.09988, 2016.

\bibitem{seshadri.bookchapter17}
V.~Seshadri \emph{et~al.}, ``{Simple Operations in Memory to Reduce Data
  Movement},'' in \emph{Advances in Computers, Volume 106}, 2017.

\bibitem{seshadri.bookchapter17.arxiv}
V.~Seshadri \emph{et~al.}, ``{The Processing Using Memory Paradigm: In-DRAM
  Bulk Copy, Initialization, Bitwise AND and OR},'' arXiv:1610.09603, 2016.

\bibitem{guo2014wondp}
Q.~Guo \emph{et~al.}, ``{3D-Stacked Memory-Side Acceleration: Accelerator and
  System Design},'' in \emph{WoNDP}, 2014.

\bibitem{cho2020mcdram}
S.~Cho \emph{et~al.}, ``{McDRAM v2: In-Dynamic Random Access Memory Systolic
  Array Accelerator to Address the Large Model Problem in Deep Neural Networks
  on the Edge},'' \emph{IEEE Access}, 2020.

\bibitem{ferreira2021pluto}
J.~D. Ferreira \emph{et~al.}, ``{pLUTo: In-DRAM Lookup Tables to Enable
  Massively Parallel General-Purpose Computation},'' arXiv:2104.07699, 2021.

\bibitem{kang1999flexram}
Y.~Kang \emph{et~al.}, ``{FlexRAM: Toward an Advanced Intelligent Memory
  System},'' in \emph{ICCD}, 1999.

\bibitem{giannoula2022sigmetrics}
C.~Giannoula \emph{et~al.}, ``{Towards Efficient Sparse Matrix Vector
  Multiplication on Real Processing-in-Memory Architectures},'' in
  \emph{SIGMETRICS}, 2022.

\bibitem{murphy2001characterization}
R.~C. Murphy \emph{et~al.}, ``{The Characterization of Data Intensive Memory
  Workloads on Distributed PIM Systems},'' in \emph{Intelligent Memory
  Systems}.\hskip 1em plus 0.5em minus 0.4em\relax Springer, 2001.

\bibitem{landgraf2021combining}
J.~Landgraf \emph{et~al.}, ``{Combining Emulation and Simulation to Evaluate a
  Near Memory Key/Value Lookup Accelerator},'' arXiv:2015.06594, 2021.

\bibitem{ahmed2019compiler}
H.~Ahmed \emph{et~al.}, ``{A Compiler for Automatic Selection of Suitable
  Processing-in-Memory Instructions},'' in \emph{DATE}, 2019.

\bibitem{singh2021fpga}
G.~Singh \emph{et~al.}, ``{FPGA-based Near-Memory Acceleration of Modern
  Data-Intensive Applications},'' \emph{IEEE Micro}, 2021.

\bibitem{rodrigues2016scattergather}
A.~Rodrigues \emph{et~al.}, ``{Towards a Scatter-Gather Architecture: Hardware
  and Software Issues},'' in \emph{MEMSYS}, 2019.

\bibitem{ghiasi2022genstore}
N.~M. Ghiasi \emph{et~al.}, ``{GenStore: A High-Performance and
  Energy-Efficient In-Storage Computing System for Genome Sequence Analysis},''
  in \emph{ASPLOS}, 2022.

\bibitem{deoliveira2021IEEE}
G.~F. Oliveira \emph{et~al.}, ``{DAMOV: A New Methodology and Benchmark Suite
  for Evaluating Data Movement Bottlenecks},'' \emph{IEEE Access}, 2021.

\bibitem{jain2018computing}
S.~Jain \emph{et~al.}, ``{Computing-in-Memory with Spintronics},'' in
  \emph{DATE}, 2018.

\bibitem{boroumand2021google_arxiv}
A.~Boroumand \emph{et~al.}, ``{Google Neural Network Models for Edge Devices:
  Analyzing and Mitigating Machine Learning Inference Bottlenecks},''
  arXiv:2109.14320, 2021.

\bibitem{Zois2018}
V.~Zois \emph{et~al.}, ``{Massively Parallel Skyline Computation for
  Processing-in-Memory Architectures},'' in \emph{PACT}, 2018.

\bibitem{boroumand2021google}
A.~Boroumand \emph{et~al.}, ``{Google Neural Network Models for Edge Devices:
  Analyzing and Mitigating Machine Learning Inference Bottlenecks},'' in
  \emph{PACT}, 2021.

\bibitem{Kautz1969}
W.~H. {Kautz}, ``{Cellular Logic-in-Memory Arrays},'' \emph{IEEE TC}, 1969.

\bibitem{nair2015evolution}
R.~Nair, ``{Evolution of Memory Architecture},'' \emph{Proceedings of the
  IEEE}, 2015.

\bibitem{lloyd2017keyvalue}
S.~Lloyd \emph{et~al.}, ``{Near Memory Key/Value Lookup Acceleration},'' in
  \emph{MEMSYS}, 2017.

\bibitem{santos2017operand}
P.~C. Santos \emph{et~al.}, ``{Operand Size Reconfiguration for Big Data
  Processing in Memory},'' in \emph{DATE}, 2017.

\bibitem{zha2020hyper}
Y.~Zha \emph{et~al.}, ``{Hyper-AP: Enhancing Associative Processing Through A
  Full-Stack Optimization},'' in \emph{ISCA}, 2020.

\bibitem{gu2020ipim}
P.~Gu \emph{et~al.}, ``{iPIM: Programmable In-Memory Image Processing
  Accelerator using Near-Bank Architecture},'' in \emph{ISCA}, 2020.

\bibitem{asgarifafnir}
B.~Asgari \emph{et~al.}, ``{FAFNIR: Accelerating Sparse Gathering by Using
  Efficient Near-Memory Intelligent Reduction},'' in \emph{HPCA}, 2021.

\bibitem{giannoula2022sparsep}
C.~Giannoula \emph{et~al.}, ``{SparseP: Towards Efficient Sparse Matrix Vector
  Multiplication on Real Processing-In-Memory Systems},'' arXiv:2201.05072,
  2022.

\bibitem{boroumand2021icde}
A.~Boroumand \emph{et~al.}, ``{Polynesia: Enabling Effective Hybrid
  Transactional Analytical Databases with Specialized Hardware Software
  Co-Design},'' in \emph{ICDE}, 2022.

\bibitem{xi2020memory}
Y.~Xi \emph{et~al.}, ``{In-Memory Learning With Analog Resistive Switching
  Memory: A Review and Perspective},'' \emph{Proceedings of the IEEE}, 2020.

\bibitem{balasubramonian2014near}
R.~Balasubramonian \emph{et~al.}, ``{Near-Data Processing: Insights from a
  MICRO-46 Workshop},'' \emph{IEEE Micro}, 2014.

\bibitem{herruzo2021enabling}
J.~M. Herruzo \emph{et~al.}, ``{Enabling Fast and Energy-Efficient FM-Index
  Exact Matching Using Processing-Near-Memory},'' \emph{The Journal of
  Supercomputing}, 2021.

\bibitem{jacob2016compiling}
A.~C. Jacob \emph{et~al.}, ``{Compiling for the Active Memory Cube},'' Tech.
  rep. RC25644 (WAT1612-008). IBM Research Division, Tech. Rep., 2016.

\bibitem{fujiki2018memory}
D.~Fujiki \emph{et~al.}, ``{In-Memory Data Parallel Processor},'' in
  \emph{ASPLOS}, 2018.

\bibitem{gokhale1995processing}
M.~Gokhale \emph{et~al.}, ``{Processing in Memory: The Terasys Massively
  Parallel PIM Array},'' \emph{IEEE Computer}, 1995.

\bibitem{elliott1999computational}
D.~G. Elliott \emph{et~al.}, ``{Computational RAM: Implementing Processors in
  Memory},'' \emph{IEEE Design \& Test of Computers}, 1999.

\bibitem{nair2015active}
R.~Nair \emph{et~al.}, ``{Active Memory Cube: A Processing-in-Memory
  Architecture for Exascale Systems},'' \emph{IBM JRD}, 2015.

\bibitem{lloyd2018dse}
S.~Lloyd \emph{et~al.}, ``{Design Space Exploration of Near Memory
  Accelerators},'' in \emph{MEMSYS}, 2018.

\bibitem{singh2020nero}
G.~Singh \emph{et~al.}, ``{NERO: A Near High-Bandwidth Memory Stencil
  Accelerator for Weather Prediction Modeling},'' in \emph{FPL}, 2020.

\bibitem{lavenier2020}
D.~{Lavenier} \emph{et~al.}, ``{Variant Calling Parallelization on
  Processor-in-Memory Architecture},'' in \emph{BIBM}, 2020.

\bibitem{kogge1994}
P.~M. Kogge, ``{EXECUBE - A New Architecture for Scaleable MPPs},'' in
  \emph{ICPP}, 1994.

\bibitem{shaw1981non}
D.~E. Shaw \emph{et~al.}, ``{The NON-VON Database Machine: A Brief Overview},''
  \emph{IEEE Database Eng. Bull.}, 1981.

\bibitem{Mai:2000:SMM:339647.339673}
K.~Mai \emph{et~al.}, ``{Smart Memories: A Modular Reconfigurable
  Architecture},'' in \emph{ISCA}, 2000.

\bibitem{besta2021sisa}
M.~Besta \emph{et~al.}, ``{SISA: Set-Centric Instruction Set Architecture for
  Graph Mining on Processing-in-Memory Systems},'' in \emph{MICRO}, 2021.

\bibitem{patterson1997case}
D.~Patterson \emph{et~al.}, ``{A Case for Intelligent RAM},'' \emph{IEEE
  Micro}, 1997.

\bibitem{yavits2021giraf}
L.~Yavits \emph{et~al.}, ``{GIRAF: General Purpose In-Storage Resistive
  Associative Framework},'' \emph{IEEE TPDS}, 2021.

\bibitem{boroumand2021polynesia}
A.~Boroumand \emph{et~al.}, ``{Polynesia: Enabling Effective Hybrid
  Transactional/Analytical Databases with Specialized Hardware/Software
  Co-Design},'' arXiv:2103.00798, 2021.

\bibitem{impica}
K.~Hsieh \emph{et~al.}, ``{Accelerating Pointer Chasing in 3D-Stacked Memory:
  Challenges, Mechanisms, Evaluation},'' in \emph{ICCD}, 2016.

\bibitem{stone1970logic}
H.~S. Stone, ``{A Logic-in-Memory Computer},'' \emph{IEEE TC}, 1970.

\bibitem{denzler2021casper}
A.~Denzler \emph{et~al.}, ``Casper: Accelerating stencil computation using
  near-cache processing,'' arXiv:2112.14216, 2021.

\bibitem{lloyd2015memory}
S.~Lloyd \emph{et~al.}, ``{In-memory Data Rearrangement for Irregular,
  Data-intensive Computing},'' \emph{Computer}, 2015.

\bibitem{sura2015data}
Z.~Sura \emph{et~al.}, ``{Data Access Optimization in a Processing-in-Memory
  System},'' in \emph{CF}, 2015.

\bibitem{wen2017rebooting}
W.-M. Hwu \emph{et~al.}, ``{Rebooting the Data Access Hierarchy of Computing
  Systems},'' in \emph{ICRC}, 2017.

\bibitem{chi2016prime}
P.~Chi \emph{et~al.}, ``{PRIME: A Novel Processing-In-Memory Architecture for
  Neural Network Computation In ReRAM-Based Main Memory},'' in \emph{ISCA},
  2016.

\bibitem{deoliveira2021}
G.~F. Oliveira \emph{et~al.}, ``{DAMOV: A New Methodology and Benchmark Suite
  for Evaluating Data Movement Bottlenecks},'' arXiv:2105.03725, 2021.

\bibitem{diab2022hicomb}
S.~Diab \emph{et~al.}, ``{High-throughput Pairwise Alignment with the Wavefront
  Algorithm using Processing-in-Memory},'' in \emph{HICOMB}, 2022.

\bibitem{Draper:2002:ADP:514191.514197}
J.~Draper \emph{et~al.}, ``{The Architecture of the DIVA Processing-in-Memory
  Chip},'' in \emph{SC}, 2002.

\bibitem{singh2021accelerating}
G.~Singh \emph{et~al.}, ``{Accelerating Weather Prediction using Near-Memory
  Reconfigurable Fabric},'' \emph{ACM TRETS}, 2021.

\bibitem{shin2018mcdram}
H.~Shin \emph{et~al.}, ``{McDRAM: Low latency and energy-efficient matrix
  computations in DRAM},'' \emph{IEEE TCADICS}, 2018.

\bibitem{gokhale2015rearr}
M.~Gokhale \emph{et~al.}, ``{Near Memory Data Structure Rearrangement},'' in
  \emph{MEMSYS}, 2015.

\bibitem{oskin1998active}
M.~Oskin \emph{et~al.}, ``{Active Pages: {A} Computation Model for Intelligent
  Memory},'' in \emph{ISCA}, 1998.

\bibitem{zheng2016tcam}
L.~Zheng \emph{et~al.}, ``{RRAM-based TCAMs for pattern search},'' in
  \emph{ISCAS}, 2016.

\bibitem{amiraliphd}
A.~Boroumand, ``{Practical Mechanisms for Reducing Processor-Memory Data
  Movement in Modern Workloads},'' Ph.D. dissertation, Carnegie Mellon
  University, 2020.

\bibitem{DBLP:conf/sigmod/BabarinsaI15}
O.~O. Babarinsa \emph{et~al.}, ``{JAFAR: Near-Data Processing for Databases},''
  in \emph{SIGMOD}, 2015.

\bibitem{diab2022high}
S.~Diab \emph{et~al.}, ``{High-throughput Pairwise Alignment with the Wavefront
  Algorithm using Processing-in-Memory},'' arXiv:2204.02085, 2022.

\bibitem{olgun2021quactrng}
A.~Olgun \emph{et~al.}, ``{QUAC-TRNG: High-Throughput True Random Number
  Generation Using Quadruple Row Activation in Commodity DRAM Chips},'' in
  \emph{{ISCA}}, 2021.

\bibitem{talukder2019exploiting}
B.~M.~S. {Bahar Talukder} \emph{et~al.}, ``{Exploiting DRAM Latency Variations
  for Generating True Random Numbers},'' in \emph{ICCE}, 2019.

\bibitem{malloc}
{Linux man-pages Project}, ``{malloc(3) — Linux manual page},''
  \url{https://man7.org/linux/man-pages/man3/malloc.3.html}, 2022.

\bibitem{posixmemalign}
{Linux man-pages Project}, ``{posix\_memalign(3) — Linux manual page},''
  \url{https://man7.org/linux/man-pages/man3/posix_memalign.3.html}, 2022.

\bibitem{x86-manual}
\BIBentryALTinterwordspacing
Intel, ``{Intel 64 and IA-32 Architectures Software Developer Manuals},'' 2011.
  [Online]. Available:
  \url{http://www.intel.com/content/www/us/en/processors/architectures-software-developer-manuals.html}
\BIBentrySTDinterwordspacing

\bibitem{seshadri2014dirty}
V.~Seshadri \emph{et~al.}, ``{The Dirty-Block Index},'' in \emph{ISCA}, 2014.

\bibitem{hassan2017softmc}
H.~Hassan \emph{et~al.}, ``{SoftMC: A Flexible and Practical Open-Source
  Infrastructure for Enabling Experimental DRAM Studies},'' in \emph{HPCA},
  2017.

\bibitem{gem5-gpu}
J.~Power \emph{et~al.}, ``{gem5-gpu: A Heterogeneous CPU-GPU Simulator},''
  \emph{CAL}, Jan 2015.

\bibitem{GEM5}
N.~Binkert \emph{et~al.}, ``{The gem5 Simulator},'' \emph{CAN}, 2011.

\bibitem{ramulator.github}
{SAFARI Research Group}, ``{Ramulator: A DRAM Simulator -- GitHub
  Repository},'' \url{https://github.com/CMU-SAFARI/ramulator/}, 2015.

\bibitem{ramulator}
Y.~Kim \emph{et~al.}, ``{Ramulator: A Fast and Extensible DRAM Simulator},''
  \emph{CAL}, 2015.

\bibitem{ramulator-pim}
S.~R. Group, ``{Ramulator-PIM: A Processing-in-Memory Simulation Framework --
  GitHub Repository},'' \url{https://github.com/CMU-SAFARI/ramulator-pim},
  2021.

\bibitem{zsim}
D.~Sanchez \emph{et~al.}, ``{ZSim: Fast and Accurate Microarchitectural
  Simulation of Thousand-Core Systems},'' in \emph{ISCA}, 2013.

\bibitem{lee.hpca15}
D.~Lee \emph{et~al.}, ``{Adaptive-Latency DRAM: Optimizing DRAM Timing for the
  Common-Case},'' in \emph{HPCA}, 2015.

\bibitem{kim2018solar}
J.~{Kim} \emph{et~al.}, ``{Solar-DRAM: Reducing DRAM Access Latency by
  Exploiting the Variation in Local Bitlines},'' in \emph{ICCD}, 2018.

\bibitem{cojocar2020susceptible}
L.~Cojocar \emph{et~al.}, ``{Are We Susceptible to Rowhammer? An End-to-End
  Methodology for Cloud Providers},'' in \emph{S\&P}, 2020.

\bibitem{salp}
Y.~Kim \emph{et~al.}, ``{A Case for Exploiting Subarray-Level Parallelism
  (SALP) in DRAM},'' in \emph{ISCA}, 2012.

\bibitem{donghyuk-ddma}
D.~Lee \emph{et~al.}, ``{Decoupled Direct Memory Access: Isolating CPU and IO
  Traffic by Leveraging a Dual-Data-Port DRAM},'' in \emph{PACT}, 2015.

\bibitem{chang.sigmetrics17}
K.~K. Chang \emph{et~al.}, ``{Understanding Reduced-Voltage Operation in Modern
  DRAM Devices: Experimental Characterization, Analysis, and Mechanisms},'' in
  \emph{SIGMETRICS}, 2017.

\bibitem{ghose2018vampire}
S.~Ghose \emph{et~al.}, ``{What Your DRAM Power Models Are Not Telling You:
  Lessons from a Detailed Experimental Study},'' in \emph{SIGMETRICS}, 2018.

\bibitem{patel2017reaper}
M.~Patel \emph{et~al.}, ``{The Reach Profiler (REAPER): Enabling the Mitigation
  of DRAM Retention Failures via Profiling at Aggressive Conditions},'' in
  \emph{ISCA}, 2017.

\bibitem{luo2020clr}
H.~Luo \emph{et~al.}, ``{CLR-DRAM: A Low-Cost DRAM Architecture Enabling
  Dynamic Capacity-Latency Trade-Off},'' in \emph{ISCA}, 2020.

\bibitem{ghose2019demystifying}
S.~Ghose \emph{et~al.}, ``{Demystifying Complex Workload-DRAM Interactions: An
  Experimental Study},'' in \emph{SIGMETRICS}, 2019.

\bibitem{kevinchang-thesis}
K.~K. Chang, ``{Understanding and Improving the Latency of DRAM-Based Memory
  Systems},'' Ph.D. dissertation, Carnegie Mellon University, 2017.

\bibitem{yoongu-thesis}
Y.~Kim, ``{Architectural Techniques to Enhance DRAM Scaling},'' Ph.D.
  dissertation, Carnegie Mellon University, 2015.

\bibitem{lee.thesis16}
D.~Lee, ``{Reducing DRAM Latency at Low Cost by Exploiting Heterogeneity},''
  Ph.D. dissertation, Carnegie Mellon University, 2016.

\bibitem{pidramarxiv}
A.~Olgun \emph{et~al.}, ``{PiDRAM: A Holistic End-to-end FPGA-based Framework
  for Processing-in-DRAM},'' arXiv:2111.00082, 2021.

\bibitem{zc706}
\BIBentryALTinterwordspacing
Xilinx, ``{Xilinx Zynq-7000 SoC ZC706 Evaluation Kit},'' 2021. [Online].
  Available:
  \url{https://www.xilinx.com/products/boards-and-kits/ek-z7-zc706-g.html}
\BIBentrySTDinterwordspacing

\bibitem{asanovic2016rocket}
K.~Asanovi{\'c} \emph{et~al.}, ``{The Rocket Chip Generator},'' Technical
  Report No. UCB/EECS-2016-17, 2016.

\bibitem{riscv-pk}
RISC-V, ``{RISC-V Proxy Kernel},''
  \url{https://github.com/riscv-software-src/riscv-pk}, 2022.

\bibitem{memcpy}
{Linux man-pages Project}, ``{memcpy(3) — Linux manual page},''
  \url{https://man7.org/linux/man-pages/man3/memcpy.3.html}, 2022.

\bibitem{calloc}
{Linux man-pages Project}, ``{calloc(3p) — Linux manual page},''
  \url{https://man7.org/linux/man-pages/man3/calloc.3p.html}, 2022.

\bibitem{youtube}
A.~Olgun, ``{Projects and Seminars on PIM, Invited Talk, End-to-end Framework
  for Processing-using-Memory},'' \url{https://youtu.be/s_z_S6FYpC8}, 2021.

\end{thebibliography}

\end{document}